\documentclass[12pt]{article}
\usepackage{putex}
\usepackage{graphicx}
\usepackage{latexsym,amsmath,amsfonts,amssymb,cite}
\usepackage{bbm}
\numberwithin{equation}{section}

\def\btab{\begin{table}[h] \begin{center} \begin{tabular}{l lp{3in}}}
      \def\etab{\end{tabular} \end{center} \end{table}}
\def\btabm{\begin{center} \begin{tabular}}
    \def\etabm{\end{tabular} \end{center}}
\def\eg{{\it e.g.}}

\def\ie{{\it i.e.}}

\def\D{{\Delta}}

\def\CD{{\cal D}}

\def\CR{{\cal R}}

\def\CN{{\cal N}}
\def\CO{{\cal O}}

\def\CW{{\cal W}}

\def\BC{\mathbb{C}}

\newcommand{\vev}[1]{ \left\langle {#1} \right\rangle }
\DeclareMathOperator{\Tr}{Tr}


\begin{document}
%
%%%%%%%% title page %%%%%%%%%
\title{On RG Flow of $\tau_{RR}$ for Supersymmetric Field Theories in Three-Dimensions}

\authors{Tatsuma Nishioka$^\dagger$, and Kazuya Yonekura$^\ddagger$}

\institution{??}{
${}^\dagger$Department of Physics, Princeton University, Princeton, NJ 08544, USA}

\institution{??}{
${}^\ddagger$Institute for Advanced Study, Princeton, NJ 08540, USA
}

\abstract{
The coefficient $\tau_{RR}$ of the two-point function of the superconformal $U(1)_R$ currents of $\CN=2$ SCFTs in three-dimensions is recently shown to be obtained by differentiating the partition function on a squashed three-sphere with respect to the squashing parameter.
With this method, 
we compute the $\tau_{RR}$ for $\CN=2$ Wess-Zumino models and SQCD 
numerically for small number of flavors and analytically in the large number limit.
We study the behavior of $\tau_{RR}$ under an RG flow by adding superpotentials to the theories.  
While the $\tau_{RR}$ decreases for the gauge theories, we find an $\CN=2$ Wess-Zumino model whose $\tau_{RR}$ increases along the RG flow.
Since $\tau_{RR}$ is proportional to the coefficient $C_T$ of the two-point correlation function of the stress-energy tensors for $\CN=2$ superconformal field 
theories, 
this rules out the possibility of $C_T$ being a measure of the degrees of freedom which monotonically decreases along RG flows in three-dimensions.
}

\preprint{PUPT-2440}

\date{March, 2013}

\maketitle
%%%%%% end of title page %%%%%%

%%%%%% table of contents %%%%%%
\tableofcontents
%%%%%%%%%%%%%%%%%%%%%

%%%%%%%%%%%%%%%%%%%%%%%%%%%%%%%%%%%%%%%%%%%%
\section{Introduction}
%%%%%%%%%%%%%%%%%%%%%%%%%%%%%%%%%%%%%%%%%%%%
Three-dimensional field theories have attracted renewed attentions since the development of localization of supersymmetric gauge theories on a three-sphere \cite{Kapustin:2009kz,Jafferis:2010un,Hama:2010av}.
One of the intriguing applications of the localization technique is 
the $F$-theorem proposed in \cite{Jafferis:2011zi} based on the extremization principle of the partition function ($Z$-extremization) \cite{Jafferis:2010un}, stating that the free energy defined by the logarithm of the partition function on $S^3$, $F=-\log Z$, decreases along any renormalization group (RG) trajectory.
For $\CN=2$ superconformal field theories, the free energy $F$ is proved to be maximized in \cite{Closset:2012vg} that establishes the $F$-theorem as long as there is no accidental symmetry in the infrared (IR) fixed point of RG flows for the same reason as the $a$-maximization in four-dimensions \cite{Intriligator:2003jj}.
The $F$-theorem was further conjectured to hold for any unitary field theory even without supersymmetry \cite{Klebanov:2011gs}.
A proof is presented in \cite{Casini:2012ei} through the relation between $F$ and the entanglement entropy of a circle for CFTs \cite{Casini:2011kv}.\footnote{Entanglement entropy of a circle and $F$ are equivalent up to ultraviolet (UV) divergent terms. One can define a renormalized entanglement entropy that interpolates the free energies at the UV and IR fixed points \cite{Liu:2012ee}, while it is not necessarily stationary against perturbation around the fixed point \cite{Klebanov:2012va} different from the Zamolodchikov's $c$-function.}
This is the analogue of the $c$-theorem in 2D \cite{Zamolodchikov:1986gt} and the $a$-theorem in 4D\cite{Cardy:1988cwa,Komargodski:2011vj} where 
the central charges are uniquely defined by the $a$-anomaly as a coefficient of the Euler characteristic in the trace of the stress-energy tensor.

It is, however, not obvious why the free energy on $S^3$ counts the number of degrees of freedom of three-dimensional field theories. 
One may well define a ``thermal central charge'' $c_\text{Therm}$ as a coefficient of the thermal free energy $F_\text{Therm} \sim c_\text{Therm} T^3$ at finite temperature $T$.
While $c_\text{Therm}$ is equivalent to the central charge for CFT$_2$ up to a constant, it can not be a $c$-function in higher dimensions. 
Indeed, $c_\text{Therm}$ increases along the RG flow from the critical $O(N)$ vector model to the Goldstone phase with spontaneously broken $O(N)$ symmetry in three-dimensions \cite{Sachdev:1993pr}.
Another possibility is the use of the coefficient $C_T$ of the two-point function of the stress-energy tensors as we will describe shortly.

The forms of the two-point functions of vector currents $J_I^\mu$ and stress-energy tensors $T_{\mu\nu}$ of conformal field theories are fixed by conformal symmetry and the conservation laws in $d$-dimensions  (see \eg \cite{Osborn:1993cr})
\begin{align}
	\langle J^\mu_I(x) J^\nu_J(0) \rangle &= \,C_{IJ}\, \frac{I_{\mu\nu} (x)}{x^{2(d-1)}} \ , \\
	\langle T_{\mu\nu}(x) T_{\rho \sigma}(0) \rangle &= \,C_T\, \frac{I_{\mu\nu, \rho\sigma} (x)}{x^{2d}} \ ,
\end{align}
where the Greek indices label the types of the currents and $\mu,\nu$ are the spacetime indices running from $1$ to $d$.
The functions $I_{\mu\nu}$ and $I_{\mu\nu,\rho\sigma}$ are defined as follows:
\begin{align}
	I_{\mu\nu}(x) &= \delta_{\mu\nu} - 2\frac{x_\mu x_\nu}{x^2} \ , \nonumber \\
	I_{\mu\nu,\rho\sigma}(x)&= \frac{1}{2}\left(I_{\mu\nu}(x) I_{\rho\sigma}(x) + I_{\mu\rho}(x) I_{\nu\sigma}(x)\right) - \frac{\delta_{\mu\nu} \delta_{\rho\sigma}}{d} \ .
\end{align}
The correlation functions are characterized by the positive coefficients $C_{IJ}$ and $C_T$ for unitary field theories.
$C_T$ is the central charge $c$ in two-dimensions, while
it is not the $a$-, but the $c$-anomaly in four-dimensions \cite{Osborn:1993cr} 
which does not necessarily decrease under RG flows \cite{Cappelli:1990yc,Anselmi:1997am}.

Less is known about $C_T$ in three-dimensions, except for the $O(N)$ vector model in the large-$N$ limit \cite{Petkou:1994ad}.
There are $N$ independent free scalar fields in the UV fixed point, leading to $C_T^\text{free} = N$, while $N-1$ fields contribute in the Goldstone phase, $C_T^\text{Goldstone}=N-1.$\footnote{Here we normalized $C_T$ so that one real free scalar field gives $C_T=1$.}
In between the RG flow, there is the critical $O(N)$ fixed point with $C_T^\text{critical}=N - \frac{40}{9\pi^2}$.
As opposed to $c_\text{Therm}$, $C_T$ decreases for the $O(N)$ vector model under the RG flow: $C_T^\text{Goldstone}<  C_T^\text{critical} < C_T^\text{free}$.
It follows from this observation that $C_T$ is conjectured to be a measure of degrees of freedom in three-dimensions \cite{Petkou:1995vu}.
Although there are no other examples nor a proof for the conjectured $C_T$-theorem in field theories,
holographic analysis of \cite{Myers:2010tj} implies its monotonicity along the RG flow in $d=3$ (and of course $d=2$) dimensions,
assuming the null energy condition and the absence of ghosts of the gravity.

In this paper, we would like to test this conjecture with various examples. Especially, we will consider $\CN = 2$ supersymmetric field theories in three-dimensions.
These theories have the $R$-symmetry current associated to the superconformal $U(1)_R$ symmetry, and 
$C_T$ is proportional to the coefficient $\tau_{RR}$ of the two-point function of the $R$-symmetry currents
\begin{align}\label{N2CF}
	\langle J^\mu_R(x) J^\nu_R(0) \rangle &= \,\frac{\tau_{RR}}{4\pi^2}\, \frac{I_{\mu\nu} (x)}{x^{4}} \ , \nonumber \\
	\langle T_{\mu\nu}(x) T_{\rho \sigma}(0) \rangle &= \,\frac{3\,\tau_{RR}}{2\pi^2}\, \frac{I_{\mu\nu, \rho\sigma} (x)}{x^{6}} \ .
\end{align}
In general, the $U(1)_R$ symmetry can mix with other global $U(1)$ symmetries, and 
one needs to determine the correct $R$-charge at the IR fixed point to compute $\tau_{RR}$ or $C_T$. 
It can be implemented by minimizing 
$\tau_{RR}$ with respect to a trial $R$-charge as shown in \cite{Barnes:2005bm}.
This, however, is not practical because
a trial $\tau_{RR}$ function has not been known for interacting field theories.
A more practical way is to employ the $Z$-extremization \cite{Jafferis:2010un,Closset:2012vg} where the partition function $Z$ on $S^3$ as a function of trial $R$-charges is minimized.
Moreover, $\tau_{RR}$ is able to be obtained by placing $\CN=2$ theories on a squashed three-sphere and taking the second derivative of the partition function with respect to the squashing parameter \cite{Closset:2012ru} as we will review in section \ref{sc:Localization}. 
So to compute $\tau_{RR}$ of $\CN=2$ theories, we will carry out the following:
1) Compute the $R$-charge $\D$ by minimizing the sphere partition function $Z$ as a function of $\D$, 2) Substitute $\D$ obtained in 1) into the second derivative of the partition function given by Eq.\,\eqref{N2tauRR}.

A class of the theories we will consider is $\CN=2$ supersymmetric QCD (SQCD) and Wess-Zumino models.
In section \ref{sc:SQCD}, we start with calculating the $\tau_{RR}$ for the non-chiral $\CN=2$ SQCD with $N_f$ flavors without superpotentials. 
The partition function is given by a multiple integral of hyperbolic gamma functions whose arguments depend on the matter contents of the theory. 
While the integral is to be performed analytically for some simple cases, it is generally beyond our ability.
Nevertheless, we are able to obtain the few leading terms of $\D$ and $\tau_{RR}$ in the large-$N_f$ limit.
The value of $\D$ is extensively studied in the context of the $F$-theorem in \cite{Klebanov:2011td,Safdi:2012re}, and we will follow their method.
Given the correct $R$-charge, a similar calculation leads us to $\tau_{RR}$ at the IR fixed point.
We study the behavior of $\tau_{RR}$ by adding a superpotential to the theory that flows to an $\CN=3$ fixed point. 
Comparing the values at the fixed points with and without the superpotential, we will show $\tau_{RR}$ decreases along the RG flow.
To tackle the small $N_f$ region, we perform numerical computations of the multiple integral. 
They fit the analytic large-$N_f$ results very well even for small $N_f$.
We argue that our large-$N_f$ results are valid in general and the $C_T$-theorem holds in these RG flows.
The chiral case is also studied with the large number of flavors, followed by the same conclusion. 
More general class of theories is investigated and the details are presented in appendix \ref{ap:B}.

The $\tau_{RR}$ decreases for the gauge theories we have considered. However, it is not the case 
with an $\CN=2$ Wess-Zumino model as we will see in section \ref{sc:WZ}.
We will consider a theory of $N+1$ chiral fields, $X$ and $Z_i\, (i=1,\dots,N)$, interacting through a superpotential of the form $\CW\sim X \sum_{i=1}^N Z_i^2$.
We let it flow to another fixed point by decoupling the $X$ field with the additional superpotential $\D\CW \sim m X^2$.
The resulting IR theory will have a quartic superpotential of $Z_i$, $\CW_\text{IR}\sim (\sum_{i=1}^N Z_i^2)^2$.
This model is reminiscent of the $O(N)$ vector model that disproves the ``$c_\text{Therm}$"-theorem. 
We confirm that this flow is consistent with the $F$-theorem (the free energy on $S^3$ decreases).
On the other hand, we find $\tau_{RR}$ increases along the RG flow and 
this model rules out the possibility of the $C_T$-theorem.

%%%%%%%%%%%%%%%%%%%%%%%%%%%%%%%%%%%%%%%%%%%%
\section{$\tau_{RR}$ in $\CN = 2$ supersymmeric field theories}\label{sc:Localization}
%%%%%%%%%%%%%%%%%%%%%%%%%%%%%%%%%%%%%%%%%%%%
Three-dimensional $\CN = 2$ superconformal field theories have $U(1)$ $R$-symmetry. Since the $R$-symmetry current is in the same multiplet as the stress-energy tensor, their two-point functions are determined by the coefficient $\tau_{RR}$ defined by Eq.\,\eqref{N2CF}.
The normalization of $\tau_{RR}$ is fixed such that a free chiral multiplet has $\tau_{RR} = \frac{1}{4}$.
In \cite{Closset:2012ru}, it was shown that the flat-space two-point correlation functions \eqref{N2CF} of $\CN=2$ SCFTs can be computed using localization on a squashed three-sphere $S_b^3$
\begin{align}
	ds^2 = \frac{1}{4} \left[ d\theta^2 + \sin^2\theta d\phi^2 + \omega^2 (d\psi + \cos\theta d\phi )^2\right] \ ,
\end{align}
where the round sphere is recovered when $\omega=1$. We parametrize $\omega$ by a squashing parameter $b$ as
$\omega = (b + b^{-1})/2$.
The squashed three-sphere $S^3_b$ preserves $SU(2)\times U(1)$ symmetry and one can put $\CN=2$ theories on it with four supercharges \cite{Imamura:2011uw,Imamura:2011wg}.
Given the partition function $Z_{S^3_b}$ and the free energy $F(b) = -\log Z_{S^3_b}$, 
$\tau_{RR}$ is obtained by taking the second derivative of $F(b)$ with respect to $b$ \cite{Closset:2012ru}:
\begin{align}\label{N2tauRR}
	\tau_{RR} = \frac{2}{\pi^2}\, \text{Re}\, \frac{\partial^2 F}{\partial b^2}\bigg|_{b=1}\,  \ .
\end{align}
The above relation allows us to compute $\tau_{RR}$ of a given SCFT in terms of the partition function $F(b)$ and we can compare the values of $\tau_{RR}$ at the UV and IR fixed points.

%%%%%%%%%%%%%%%%%%%%%%%%%%%%%%%%%%%%%%%%%%%%
\subsection{Localization on a squashed three-sphere}
%%%%%%%%%%%%%%%%%%%%%%%%%%%%%%%%%%%%%%%%%%%%
The partition function of $\CN=2$ theories on $S^3_b$ are obtained via the localization \cite{Imamura:2011uw,Imamura:2011wg} in the following way\,\footnote{In the original literatures \cite{Imamura:2011uw,Imamura:2011wg}, the double sine functions are used instead of the hyperbolic gamma functions. The former is roughly the inverse of the latter. The latter is suitable for studying dualities \cite{Willett:2011gp,Benini:2011mf} that are recast as the identities between the integrals of the hyperbolic gamma functions \cite{van2007hyperbolic}.}:
\begin{itemize}
	\item The one-loop matter contribution to the partition function is given by
	\begin{align}  \label{eq:matterloop}
		Z_\text{matter}^\text{1-loop} &= \prod_I \prod_{\rho\in \CR_I} \Gamma_h \left[ \omega ( \rho(\sigma) + i \D_I)\right]  \ ,
	\end{align}
	where $\Gamma_h [z] \equiv \Gamma_h(z; i\omega_1, i\omega_2)$ is the hyperbolic gamma function defined by Eq.\,\eqref{HypGamma}. Here, we choose the special values of the arguments $\omega_1 = b$, $\omega_2 = 1/b$, and $\omega = (\omega_1 + \omega_2)/2$.
	$I$ labels the types of chiral multiplets and $\rho$ is a weight in a representation $\CR_I$. $\D_I$ is the $R$-charge of the scalar field in a chiral multiplet.
	
	\item The one-loop gauge contribution to the partition function and the path integral measure of the zero modes are combined to
	\begin{align}\label{eq:gaugeloop}
		 Z_\text{gauge}^\text{1-loop} \cdot [d\sigma ] =\frac{(2 \pi \omega)^{\text{rank}\, G}}{|W| \Vol(T)} 
		 \left( \prod_{\alpha} \Gamma_h\left[  \omega \alpha(\sigma) \right] \right)^{-1} \cdot \prod_{i=1}^{\text{rank}\, G} d\sigma_i    \ ,
	\end{align}
	where $|W|$ is the order of the Weyl group $W$ of the group $G$, $\Vol(T)$ is the volume of the maximal torus $T$ of $G$ 
	(\eg, $T=U(1)^N$ for $G=U(N)$), and $\alpha$ is a root of $G$.  
	The factor $(2 \pi \omega)^{\text{rank}\, G}$ comes from the one-loop determinant of the 
	gauge multiplets in the Cartan subalgebra. The $\omega$ dependence of this factor is important for calculations of $\tau_{RR}$.
	See appendix \ref{ap:B} for the form of this one-loop determinant without gauge-fixing.
	
	\item The Chern-Simons term of level $k$ and FI term of parameter $\xi$ give a classical contribution
	\begin{align}
		Z_\text{cl} = \exp \left[ -\pi i k \omega^2 \text{Tr}(\sigma^2) - 2\pi i \xi \omega \text{Tr} (\sigma) \right] \ .
	\end{align}
	We assume that the normalization of $\Tr$ is chosen such that the Chern-Simons level $k$ is quantized to be an integer. 
	In the case $G=U(N)$,  $\Tr$ is just the trace in the fundamental representation, $\Tr=\Tr_f$. Throughout this paper we will drop off the FI term.	
	\end{itemize}
We shall use these formulae to compute $\tau_{RR}$ for several examples below. 

%%%%%%%%%%%%%%%%%%%%%%%%%%%%%%%%%%%%%%%%%%%%
\subsection{Calculation of $\tau_{RR}$-coefficient}
%%%%%%%%%%%%%%%%%%%%%%%%%%%%%%%%%%%%%%%%%%%%
\paragraph{Wess-Zumino model:}
First, we consider a free chiral multiplet with the $R$-charge $\D $. In this case, 
the one-loop partition function \eqref{eq:matterloop} gives the exact answer. 
The integral representation of the hyperbolic gamma function (\ref{HypGamma}) is useful to see the $b$-dependence of the free energy 
\begin{align}
	F_\text{chiral}(b) = - \int_0^\infty \frac{dx}{2x}\left( \frac{\sinh (2(1-\Delta) \omega x) }{\sinh(b x) \sinh (b^{-1} x)} - \frac{2(1-\Delta)
	\omega}{x}\right) \ .
\end{align}
The $\tau_{RR}$ given by Eq.\,\eqref{N2tauRR} leads
\begin{align}
\tau_{RR}(\Delta)=&\frac{2}{\pi^2}  \int^\infty_0 dx \Bigg[ 
\left(1-\Delta \right) \left( \frac{1}{x^2}-\frac{\cosh (2x(1-\Delta))}{\sinh^2 (x)} \right) 
+ \frac{(\sinh (2x)-2x) \sinh(2x(1-\Delta))}{2\sinh^4(x)}    \Bigg] \ .\nonumber 
\end{align}
For a free chiral multiplet with $\Delta=1/2$, a short computation yields $\tau_{RR} = \frac{1}{4}$.
One can also check that $\tau_{RR}(\Delta)$ is not extremized at $\Delta=1/2$. Therefore this example excludes the possibility that
$\tau_{RR}(\Delta)$ defined by Eq.\,\eqref{N2tauRR} is extremized at the correct $R$-charges.

As a slightly nontrivial example, let us consider a Wess-Zumino model which consists of a chiral field $X$ with a cubic superpotential,
\begin{align}
\CW = X^3 \ .
\end{align}
This model has an interacting IR fixed point with $\Delta_{\rm IR}=2/3$ and $\tau_{RR}^{\rm IR}=\tau_{RR}(2/3) \simeq 0.182$. 
The UV limit is just a free chiral multiplet and hence $\tau_{RR}^{\rm UV}=1/4$.
In this model, $\tau_{RR}$ decreases along the RG flow, $\tau_{RR}^{\rm IR}<\tau_{RR}^{\rm UV}$.

\paragraph{Chern-Simons theory:}
We next consider $U(N)_k$ pure Chern-Simons theory whose partition function is 
\begin{align}
	Z_\text{CS} = \frac{1}{N!} \int \prod_{i=1}^N \omega\,d\sigma_i \, e^{-\pi i k \omega^2 \sum_{i=1}^N \sigma_i^2} 
	\prod_{\alpha \in \Delta_+}4\sinh\left( \pi b \omega \alpha(\sigma) \right) \sinh\left( \pi b^{-1} \omega \alpha(\sigma) \right) \ ,
\end{align}
where we used the identity \eqref{HGID}, and $\alpha$ runs over the positive roots of $U(N)$. 
We rewrite the partition function with the Weyl denominator formula and perform the multiple integrals to get (see \eg \cite{Kapustin:2009kz})
\begin{align}\label{eq:CSpartition}
	Z_\text{CS} = \frac{\exp\left[ \frac{\pi i}{6k} N(N^2 - 1) (1-2\omega^2) \right]}{k^{N/2}} \prod_{m=1}^{N-1}\left( 2\sin\left( \frac{\pi m}{k} \right) \right)^{N-m} \ .
\end{align}
Since the $\omega$-($b$-)dependence appears only in the imaginary part of the free energy, $\tau_{RR}$ vanishes for the CS theory, $\tau_{RR}|_{CS} = 0$. This is consistent with the fact that topological theories can not have non-zero $C_T$-coefficient since they do not couple to the background metric.\footnote{This result also gives a consistency check on the $\omega$ dependence of the overall factor of the partition function 
(\ref{eq:gaugeloop}), 
which was neglected in \cite{Imamura:2011wg}.}

\paragraph{Large-$N$ gauge theories:}
In a certain large-$N$ limit of $\CN=2$ SCFTs, the dependence on the squashing parameter $b$ of the partition function becomes 
very simple \cite{Imamura:2011wg} (see also \cite{Martelli:2011fw})
\begin{align}\label{LargeNF}
	F(b) = \frac{(b+b^{-1})^2}{4} F(1) \ ,
\end{align}
where $F(1)$ is the round three-sphere partition function.
Combining Eqs.\,\eqref{N2tauRR} and \eqref{LargeNF}, we obtain the universal result for $\tau_{RR}$
\begin{align}
	\tau_{RR} = \frac{4}{\pi^2} \,F(1) \ .
\end{align}
This is consistent with the holographic analysis in \cite{Barnes:2005bw}.
Since $C_T \propto \tau_{RR}$ in $\CN=2$ SCFTs, it follows that $C_T$ is proportional to the round sphere partition function
\begin{align}
	C_T = \frac{6}{\pi^4} \, F(1) \ .
\end{align}
The $F$-theorem \cite{Jafferis:2011zi,Casini:2012ei} ensures $F(1)$ decreases under any RG flow. 
Then the above relation between $C_T$ and $F(1)$ indicates $C_T^\text{IR} < C_T^\text{UV}$ for $\CN=2$ SCFTs in the large-$N$ limit.

%%%%%%%%%%%%%%%%%%%%%%%%%%%%%%%%%%%%%%%%%%%%
\section{$\CN = 2$ $U(N_c)$ gauge theory with flavors}\label{sc:SQCD}
%%%%%%%%%%%%%%%%%%%%%%%%%%%%%%%%%%%%%%%%%%%%

Consider $\CN=2$ $U(N_c)$ Chern-Simons theory of level $k$ coupled to $N_f$ quarks $Q_a$ and $\tilde N_f$ anti-quarks $\tilde Q_a$ 
in the fundamental and anti-fundamental representations of $U(N_c)$, respectively.
This theory enjoys $U(N_f) \times U(\tilde N_f)$ global symmetries acting on the quarks $Q_a$ and the anti-quarks $\tilde Q_a$ as vector representations, that make all the $R$-charges of $Q_a$ ($\tilde Q_a$) be equal.
These symmetries leave us two $R$-charges $R[Q_a] = \D$ and $R[\tilde Q_a] = \tilde\D$.
Since the  $U(1)_F^2 \subset U(N_f) \times U(\tilde N_f)$ flavor symmetry can mix with the $R$-symmetry, the correct values of the $R$-charges 
at the IR fixed point should be fixed by minimizing the partition function with respect to $\D$ and $\tilde \D$ \cite{Jafferis:2010un,Closset:2012vg}.

Following the rules described in section \ref{sc:Localization}, one can write down the partition function of $U(N_c)_k$ chiral SQCD with flavors on $S_b^3$
\begin{align}\label{PFb}
	Z(b) = \frac{1}{N_c!} \int \prod_{i=1}^{N_c} &d\sigma_i \, e^{-i\pi k \sum_{i=1}^{N_c} \sigma_i^2}\, Z_g (b, \sigma)\, \prod_{i=1}^{N_c}  \Gamma_h [\sigma_i + i\omega \D]^{N_f} \, \Gamma_h [-\sigma_i + i\omega \D]^{\tilde N_f} \ ,
\end{align}
where we rescaled the integration variable $\sigma$ by $\omega$ and defined the function $Z_g$ as
\begin{align}
	Z_g(b,\sigma) = \prod_{i<j}4 \sinh\left( \pi b (\sigma_i - \sigma_j) \right)\, \sinh\left( \pi b^{-1} (\sigma_i - \sigma_j) \right) \ .
\end{align}
We will assume $N_f \ge \tilde N_f$ and introduce new parameters $\bar N_f$ and $\mu$:
\begin{align}
	N_f = (1+\mu)\bar N_f \ , \qquad \tilde N_f = (1-\mu) \bar N_f \ , \qquad 0\le \mu \le 1 \ .
\end{align}
We will find it convenient to use these parametrization in section \ref{ssc:Chiral} .
Using Eq.\,\eqref{N2tauRR} and the integral representation of the hyperbolic gamma function \eqref{HypGamma}, $\tau_{RR}$ of the SQCD becomes
\begin{align}\label{tauRRSQCD}
	\tau_{RR} = & \text{Re}\,\Bigg[  \frac{1}{ N_c!\, Z(1)}
	\int \prod_{i=1}^{N_c} d\sigma_i \, e^{-i\pi k \sum_{i=1}^{N_c} \sigma_i^2}\, e^{N_f (\ell (1-\D + i\sigma) + \ell (1-\D - i\sigma))} \, Z_g(1,\sigma)
	\nonumber \\
	 & \qquad \cdot \left(g(\sigma)+ \bar N_f \sum_{i=1}^{N_c} \left[ (1+\mu)f(1 - \D ,\sigma_i) + (1-\mu)f(1 - \tilde\D ,-\sigma_i) \right]   \right) \Bigg]\,  \ ,
\end{align}
where $\ell(z)$ is the Jafferis's $\ell$-function \cite{Jafferis:2010un} defined by Eq.\,\eqref{Jafferisl} and 
\begin{align}
          f(z,\sigma) &= \frac{2}{\pi^2} \int_0^\infty dx \left[
	 \frac{z}{x^2} - \frac{z\, \cosh(2x (z + i\sigma) ) }{\sinh^2(x)} + \frac{(\sinh(2x)-2x)\sinh(2x(z+i\sigma))}{2\sinh^4 (x)} \right] \ , \\
	 g(\sigma)&=-\frac{2}{\pi^2} \sum_{i<j} 
	 \frac{(\pi (\sigma_i-\sigma_j) ) \sinh (2 \pi (\sigma_i-\sigma_j)) -2(\pi (\sigma_i-\sigma_j))^2}{\sinh^2(\pi (\sigma_i-\sigma_j))} \ .
\end{align}
To compute the value of $\tau_{RR}$ at the IR fixed point, we extremize the partition function with respect to the $R$-charge and determine $\D_\text{IR}$ and $\tilde\D_\text{IR}$. 
In the case with $N_c=N_f=\tilde N_f =1$, it is analytically computed to be $\D_\text{IR}= \tilde \D_\text{IR}=1/3$ \cite{Jafferis:2010un}.
More generally, we have to rely on numerical computations to fix them \cite{Willett:2011gp}.

Before discussing the details of the IR fixed points, let us comment on the UV limit of the theory.
If we include the usual Yang-Mills action, that is irrelevant in the IR but relevant in the UV, the theory becomes free in the UV with $N_c(N_f+\tilde{N}_f)$ chiral fields and $N_c^2$ gauge supermultiplets.
An $\CN =2$ gauge multiplet has the same propagating degrees of freedom as a chiral multiplet in three-dimensions,\footnote{Here we neglect the 
topological degrees of freedom of gauge fields. Such topological degrees of freedom is irrelevant to the correlation function of the stress tensor,
but contributes to $F$.} 
so the UV theory may be considered to be a theory of free $N_c(N_f+\tilde{N}_f+N_c)$ chiral multiplets.
However, the stress tensor is not traceless unless we perform an appropriate improvement 
$T_{\mu\nu} \to T_{\mu\nu}+(\delta_{\mu\nu} \partial^2-\partial_\mu \partial_\nu) \CO$ which breaks shift symmetries of the scalars dual to gauge fields.
The two-point function of the stress tensors takes the form of
\begin{align}
\vev{T_{\mu\nu}(x)T_{\rho\sigma}(0)}=
\frac{1}{64\pi^2}\left[ (P_{\mu \rho}P_{\nu \sigma}+P_{\mu \sigma}P_{\nu \rho}-P_{\mu \nu}P_{\rho \sigma}) \frac{\tau_{RR}}{x^2}
+ P_{\mu \nu}P_{\rho \sigma} \frac{\tau'_{RR}}{x^2} \right] \ ,
\end{align}
where we have defined
$P_{\mu\nu}=\delta_{\mu\nu} \partial^2-\partial_\mu \partial_\nu$. 
It is clear that $\tau_{RR}$ is invariant under the improvement of the stress tensor,
while $\tau'_{RR}$ is not. We use this definition of $\tau_{RR}$ for the UV theory, resulting in
\begin{align}\label{tauUV}
\tau_{RR}^{\rm UV}=\frac{N_c(2 \bar N_f +N_c)}{4} \ .
\end{align}

%%%%%%%%%%%%%%%%%%%%%%%%%%%%%%%%%%%%%%%
\subsection{Non-chiral theory}
%%%%%%%%%%%%%%%%%%%%%%%%%%%%%%%%%%%%%%%
For the non-chiral theories with $\mu=0$, there is an additional charge conjugation symmetry that exchanges the roles of quarks and anti-quarks. We have only one $R$-charge to vary: $\D = \tilde \D$.

%%%%%%%%%%%%%%%%%%%%%%%%%%%%%%%%%%%%%%%%%%%%
\subsubsection{Large-$N_f$ limit}
%%%%%%%%%%%%%%%%%%%%%%%%%%%%%%%%%%%%%%%%%%%%
While the numerical computation of $\tau_{RR}$ is to be carried out without undue effort for small $N_f$ as we shall see shortly in the next section, 
it becomes intractable for large $N_f$ because the value of the partition function gets very small and we have to increase a precision of our numerics to obtain reliable values.
In the latter case, however, the large-$N_f$ expansion is a useful analytic method we can employ.
The free energies of SQCD on a three-sphere are systematically studied in this limit in \cite{Klebanov:2011td,Safdi:2012re}. 

Given an expansion of the IR $R$-charge around $N_f=\infty$
\begin{align}\label{Rexpand}
	\D_{\CN=2} = \frac{1}{2} + \frac{\D_1}{N_f} + \frac{\D_2}{N_f^2} + \cdots \ ,
\end{align}
one can minimize the partition function $Z(1)$ with respect to $\D_1$ at some order of $1/N_f$ and obtains the value of $\D_1$.
Iterating this procedure, one finds $\D_2$ and so on at a higher order of $1/N_f$.
Using the partition function \eqref{PFb} on a round three-sphere ($b=1$), we obtain the $R$-charge at the IR fixed point up to the order of $1/N_f^2$
\begin{align}
	\D_1 &= -\frac{2N_c}{\pi^2 (\kappa^2 +1)} \ , \nonumber \\
	\D_2 &= \frac{2 \left(\left(3 \left(\pi^2-4\right) \kappa^2 - 5 \pi^2 + 36 \right) N_c^2 + \pi^2 \left(-3 \kappa^4 + 3 \kappa^2 + 2\right)\right)}{3 \pi^4
   \left(\kappa^2+1\right)^3} \ ,
\end{align}
where $\kappa = \frac{2k}{\pi N_f}$.
More terms with $\kappa = 0$ are available in \cite{Safdi:2012re}.

Once the IR $R$-charge is determined, we can proceed to compute $\tau_{RR}$ 
by substituting Eq.\,\eqref{Rexpand} into Eq.\,\eqref{tauRRSQCD}.
Performing the similar procedure to the computation of the $R$-charge leads 
\begin{align}\label{tauRRNf}
	\tau_{RR}^{\CN=2} =& \frac{N_c N_f}{2} + \frac{(32 - 3\pi^2)N_c^2}{6\pi^2 (\kappa^2 +1)} \nonumber\\
	&+ \frac{N_c}{18 \pi^4
   \left(\kappa^2+1 \right)^3 N_f} \Big[2 \left(-3 \pi^4 \kappa^4+3 \left(88-58 \pi^2+3 \pi^4\right) \kappa^2+26 \pi^2-504 \right)N_c^2 \nonumber \\
   & \qquad\qquad + \pi^2 \left(12 \left(14-\pi ^2\right) \kappa^4+9 \pi^2
   \kappa^2+9 \pi^2-40\right)\Big]
    + O(1/N_f^2) \ .
\end{align}
The UV value of $\tau_{RR}$ is given by $\tau_{RR}^{\rm UV}=\frac{1}{4}N_c (2N_f + N_c)$ as discussed above and it is always greater than Eq.\,\eqref{tauRRNf} in the large-$N_f$ limit.

We have considered the $\CN=2$ cases without superpotential, but we can introduce a superpotential $\lambda (Q T^a \tilde{Q})^2$, where $T^a$ are
generators of the gauge group $U(N_c)$. 
The theory flows to the ${\cal N}=3$ fixed point with $\lambda=2 \pi /k$ in the IR limit. In this case the R-charge is given by $\Delta=1/2$ and hence
\begin{align}\label{tauRRNfN=3}
\tau_{RR}^{{\cal N}=3}&=\frac{N_cN_f}{2} + \frac{(8 - \pi^2)N_c^2}{2\pi^2(\kappa^2 +1)} \nonumber\\
+& \frac{N_c \left(-2 \left(\left(\pi^2 \left(\kappa^2-3\right)+40\right) \kappa^2+8\right)
   N_c^2+\pi ^2 \left(-4 \kappa^4+3 \kappa^2+3\right)+8 \left(6 \kappa^4+\kappa^2-1\right)\right)}{6 \pi^2 \left(\kappa^2+1\right)^3 N_f} \nonumber \\
   &\quad + O(1/N_f^2) \ .
\end{align}
This is always smaller than that for $\CN=2$ theories \eqref{tauRRNf} as long as $N_f \gg 1$.

In summary, we consider the following RG flows between the fixed points,
\begin{align*}
\text{UV free theory} \quad \to \quad  {\cal N}=2~\text{fixed point} \quad \to \quad {\cal N}=3~\text{fixed point} \ ,
\end{align*}
and see the value of $\tau_{RR}$ decreases as 
\begin{align}\label{tauRGflow}
\tau_{RR}^{\rm UV}>\tau_{RR}^{{\cal N}=2}>\tau_{RR}^{{\cal N}=3} \ .
\end{align}

%%%%%%%%%%%%%%%%%%%%%%%%%%%%%%%%%%%%%%%%%%%%
\subsubsection{Numerical computation for small $N_f$}
%%%%%%%%%%%%%%%%%%%%%%%%%%%%%%%%%%%%%%%%%%%%
We will study how $\tau_{RR}$ behaves for small $N_f$ where the previous analysis may not be valid. There is no good approximation available, and we have to evaluate the integral given by Eq.\,\eqref{tauRRSQCD} in some way.
To determine the correct IR $R$-charge, we numerically minimized the partition function \eqref{PFb} on a round sphere ($b=1$) with respect to $\D$. Then 
we performed the numerical integrations for $\tau_{RR}$ as well using $\D$ obtained by the $Z$-minimization.
Some of the values are summarized in Table\ \ref{tab:tauRR}.

\begin{table}[htb]
\centering
	\begin{tabular}{c|cccccccc
	}
		\hline
		$N_f$ & 1 & 2 & 3 & 4 & 5 & 6 & 7 & 8 
		\\
		\hline\hline
		$\D_{\CN=2}$ & 1/3 & 0.4085 & 0.4369 & 0.4519 & 0.4611 & 0.4674 & 0.4719 & 0.4753 
		\\[0.1cm]
		$\tau_{RR}^{\CN=2}$ & 0.545 & 1.042 & 1.541 & 2.040 & 2.540 & 3.040 & 3.539 & 4.039 
		\\[0.1cm]
		$\tau_{RR}^{\CN=3}$ & 0.500 & 0.956 & 1.439 & 1.931 & 2.425 & 2.922 & 3.419 & 3.918 
		\\
		\hline
	\end{tabular}
	
	\vspace{1cm}
	
	\begin{tabular}{c|cccccccc}
		\hline
		$N_f$ &  2 & 3 & 4 & 5  & 6 & 7 & 8\\
		\hline\hline
		$\D_{\CN=2}$ & 1/4 & 0.3417 & 0.3851 & 0.4101 & 0.4262 & 0.4375 & 0.4458 \\[0.1cm]
		$\tau_{RR}^{\CN=2}$ & 1.500 & 2.670 & 3.775 & 4.844 & 5.893 & 6.929 & 7.956 \\[0.1cm]
		$\tau_{RR}^{\CN=3}$ & - & 1.939 & 3.154 & 4.267 & 5.337 & 6.346 & 7.295 \\
		\hline
	\end{tabular}
	\caption{The values of $\D_{\CN=2}$, $\tau_{RR}^{\CN=2}$ and $\tau_{RR}^{\CN=3}$ for the non-chiral SQCD with $N_c=1$ [Upper] and $N_c=2$ [Lower]. }
	\label{tab:tauRR}
\end{table}

\begin{figure}[htbp]
	\centering
	\includegraphics[width=6cm]{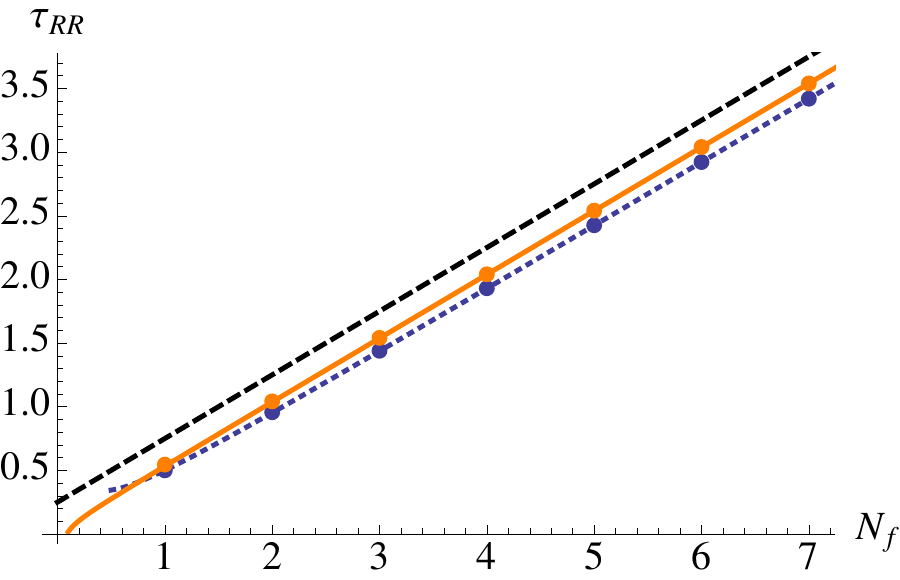}\qquad
	\includegraphics[width=6cm]{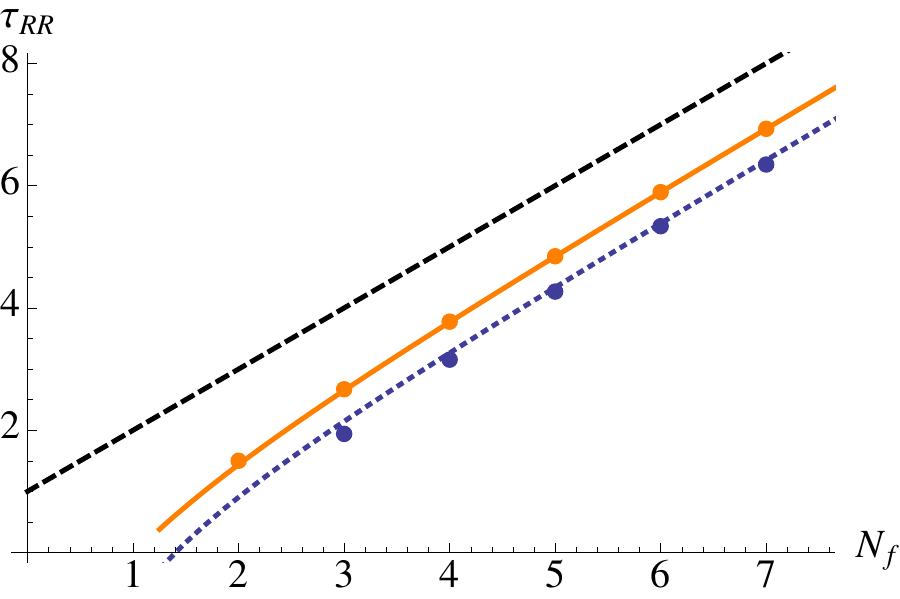}
	\caption{Plots of $\tau_{RR}^{\CN=2}$ and $\tau_{RR}^{\CN=3}$ as functions of $N_f$ for the non-chiral $U(N_c)$ SQCD of $N_c=1$ [Left] and $N_c =2$ [Right]. The solid orange and dotted blue curves are drawn using the large-$N_f$ expansion at the $\CN=2$ and $\CN=3$ fixed points, respectively. The dashed black lines are the values at the UV fixed point. The orange and blue dots are computed  numerically. They fit the large-$N_f$ approximation very well even for small $N_f$.}
	\label{fig:tauRR}
\end{figure}

Figure \ref{fig:tauRR} shows a plot of $\tau_{RR}$ as a function of $N_f$ for $N_c=1,2$ and $k=0$.\footnote{
Note that the case with $N=N_c=N_f$ needs special attention.
The $N=1$ theory flows to the $\CN=3$ (or more precisely ${\cal N}=4$ when $k=0$) IR fixed point which can be described by a free (twisted) hypermultiplet, leading to $\tau_{RR}^{\CN =3}=2\cdot \frac{1}{4}=0.500$.
For the $N=2$ theory, it flows to the $\CN=2$ free theory where six chiral fields have canonical dimension $1/2$ \cite{Aharony:1997bx}, consistent with $\tau_{RR}^{\CN=2}=6\cdot \frac{1}{4}=1.500$,
while the partition function diverges at the ${\cal N}=3$ (or $\CN=4$) fixed point. This may be related to the fact that this theory does not flow to a standard critical point~\cite{Gaiotto:2008ak} and an accidental symmetry may appear in the IR.
We would like to thank I.\,Yaakov for suggesting this possibility to us.
}
The large-$N_f$ approximation \eqref{tauRRNf} and \eqref{tauRRNfN=3} gives the orange solid and blue dotted curves for the $\CN=2$ and $\CN=3$ fixed points, respectively. The orange and blue dots are plotted numerically. They fit the large-$N_f$ approximation curves very well even for small $N_f$.
The dashed black lines are for the UV fixed point where $\tau_{RR}^{\rm UV}=N_c(2N_f+N_c)/4$.
The black, orange and blue curves are ordered as in Eq.\,\eqref{tauRGflow}.
Our results here show the validity of the large-$N_f$ analysis in the previous section even for small $N_f$.
With these observations, we argue that the monotonicity of $\tau_{RR}$ given by Eq.\,\eqref{tauRGflow} holds for arbitrary $N_c$, $N_f$ and $k$.

%%%%%%%%%%%%%%%%%%%%%%%%%%%%%%%%%%%%%%%%%%%%
\subsection{Chiral theory}\label{ssc:Chiral}
%%%%%%%%%%%%%%%%%%%%%%%%%%%%%%%%%%%%%%%%%%%%
We have treated the non-chiral case so far, but the previous discussion is easily generalized to the chiral theory. 
Now there are two independent $R$-charges $\D$ and $\tilde \D$ with respect to which the partition function is minimized.
Expanding $\D$ and $\tilde \D$ in the large-$\bar N_f$ limit and extremizing the partition function on a round sphere term by term, we find \cite{Klebanov:2011td}
\begin{align}
	\D &= \frac{1}{2} - \frac{2(1+\mu)N_c}{\pi^2 (1+\kappa^2)\bar N_f} + O(1/\bar N_f^2) \ , \nonumber \\
	\tilde \D&= \frac{1}{2} - \frac{2(1-\mu)N_c}{\pi^2 (1+\kappa^2)\bar N_f} + O(1/\bar N_f^2) \ .
\end{align}
Substituting them into Eq.\,\eqref{tauRRSQCD} leads $\tau_{RR}$ of $\CN=2$ chiral SQCD
\begin{align}\label{tauCSQCD}
	\tau_{RR} &= \frac{N_c \bar N_f}{2} + \frac{(32-3\pi^2)N_c^2}{6\pi^2 (1+\kappa^2)} + O(1/\bar N_f) \ .\end{align}
This is smaller than the UV value given by Eq.\,\eqref{tauUV}.

Introducing a superpotential $\CW \sim \lambda (Q T^a \tilde Q)^2$ lets the theory flow to another fixed point where 
the $R$-charges have to satisfy the constraint $\D + \tilde \D = 1$. Extremizing the partition function under the constraint yields the $R$-charges \cite{Klebanov:2011td}
\begin{align}
	\D &= \frac{1}{2} - \frac{4\mu N_c}{\pi^2 (1+\kappa^2)\bar N_f} + O(1/\bar N_f^2) \ , \nonumber \\
	\tilde \D&= 1-\D \ .
\end{align}
With these values, we obtain
\begin{align}
	\tau_{RR} &= \frac{N_c \bar N_f}{2} + \frac{(8(3+\mu^2)-3\pi^2)N_c^2}{6\pi^2 (1+\kappa^2)} + O(1/\bar N_f) \ , 
\end{align}
which is less than Eq.\,\eqref{tauCSQCD} for any $\mu$ between $0$ and $1$.
So even in the chiral theory, $\tau_{RR}$ decreases along RG flows.

%%%%%%%%%%%%%%%%%%%%%%%%%%%%%%%%%%%%%%%%%%%%
\subsection{More general theories}
%%%%%%%%%%%%%%%%%%%%%%%%%%%%%%%%%%%%%%%%%%%%
We can extend the large-$N_f$ calculation to general gauge groups $G=\otimes_{A} G_A$ and general representations 
of matter fields $\CR_I=\otimes_{A} \CR_{I,A}$,
where $A$ labels each simple or $U(1)$ gauge group $G_A$. 
Here we have no superpotential, and impose technical assumptions on $U(1)$ charges given by Eqs.\,(\ref{eq:cond1}) and (\ref{eq:cond2})
(the assumption Eq.\,(\ref{eq:cond1}) is not satisfied in the chiral $U(N)$ model discussed in the previous subsection).
The details are described in appendix \ref{ap:B} and we sketch the results below.
$\tau_{RR}$ is generally given by
\begin{align}
\tau_{RR}=\frac{1}{4}\left [ N_{\rm total}+\frac{2(32-3\pi^2)}{ 3\pi^2}  \sum_A \frac{\dim G_A}{1+\kappa_A^2} \right] + O(1/N_f) \ ,
\end{align}
where $N_{\rm total}$ is the total number of chiral matter fields (which is given by $N_{\rm total}=\sum_I N_I \dim \CR_I$ in the notation of appendix~\ref{ap:B}), 
and $\kappa_A $ is defined in Eq.\,(\ref{eq:kappaA}).
This should be compared with the UV value
\begin{align}
\tau_{RR}^{\rm UV}=\frac{1}{4}\left [ N_{\rm total}+  \sum_A \dim G_A \right]   \ .
\end{align}
The leading contribution to $\tau_{RR}$ in the large-$N_f$ limit is the same in the UV and IR, and the difference appears in the coefficient of $\dim G_A$ 
in the subleading term. One  can see that $\tau_{RR}$ is also decreasing in this class of theories.

%%%%%%%%%%%%%%%%%%%%%%%%%%%%%%%%%%%%%%%%%%%%
\section{RG flow with increasing $\tau_{RR}$}\label{sc:WZ}
%%%%%%%%%%%%%%%%%%%%%%%%%%%%%%%%%%%%%%%%%%%%

We have considered the RG flow of the gauge theories where $\tau_{RR}$ monotonically decreases so far.
In this section, however, a Wess-Zumino model we will discuss
has an RG flow where $\tau_{RR}$ increases. 

The model consists of $N+1$ chiral fields denoted by $X$ and $Z_i~(i=1,\cdots,N)$, and we do not introduce any gauge field. 
We assume that $Z_i$ are in the fundamental representation
of a global $O(N)$ symmetry, that allows the following superpotential:
\begin{align}
\CW=X \sum_{i=1}^N (Z_i)^2  \ .  \label{eq:WZsuperpot}
\end{align}
This model has an interacting fixed point due to the superpotential.

In the large-$N$ limit, it is easy to obtain $\tau_{RR}$ analytically. First, the $R$-charges of $Z_i$ and $X$, denoted as $\Delta_Z$ and
$\Delta_X$ respectively, are obtained by extremizing 
$F=- N\ell(1-\Delta_Z)-\ell(1-\Delta_X)$ under the constraint $2\Delta_Z+\Delta_X=2$. After a short calculation, we obtain
\begin{align}
\Delta_Z &=\frac{1}{2}+\frac{4}{\pi^2 N}+\frac{32}{\pi^4 N^2}+O(1/N^3)  \ ,  \\
\Delta_X &= 2(1-\Delta_Z) \ .
\end{align}
Using these $R$-charges, the $\tau_{RR}$ and the free energy $F$ are calculated in the same way as in the previous section
\begin{align}
\tau_{RR}^{XZ} &=\frac{N}{4}- \frac{4}{3\pi^2 }+\left( \frac{68}{9 \pi^2}-\frac{48}{\pi^4}  \right) \frac{1}{N}+O(1/N^2)  \ , \label{eq:WZlargeN} \\
F^{XZ} &=\frac{N}{2} \log 2 +\frac{4}{\pi^2 N} +O(1/N^2) \ . \label{WZlargeNF}
\end{align}
These $\tau_{RR}$ and $F$ are smaller than the value of the UV fixed point $\tau_{RR}^\text{free} = \frac{N+1}{4}$ and $F^\text{free}=\frac{N+1}{2} \log 2$
where one can neglect the superpotential and there are $N+1$ free chiral fields.

Next, let us add a mass term to $X$,
\begin{align}
\D \CW= mX^2 \ .
\end{align}
Integrating out $X$ leads to the theory of $N$ chiral multiplets $Z_i$ with 
the quadratic superpotential $\CW_\text{IR} \sim (\sum_{i=1}^N (Z_i)^2)^2$.
This superpotential is marginally irrelevant and hence the IR theory is just a free theory of $N$ chiral multiplets.
The $R$-charge of $Z$ is $\D_Z = \frac{1}{2}$,
and the $\tau_{RR}$ and $F$ at the IR fixed point is equal to that of $N$ free chiral multiplets:
\begin{align}
\tau_{RR}^{Z}&=\frac{N}{4}  \ , \label{WZIR}\\
F^{Z}&=\frac{N}{2} \log 2 \ .\label{WZIRF}
\end{align}
Comparing Eq.\,\eqref{eq:WZlargeN} and Eq.\,\eqref{WZIR}, 
we find that $\tau_{RR}$ increases for sufficiently large-$N$ under the RG flow, while the free energy $F$ is decreasing as expected from the 
$F$-theorem.

\begin{figure}[htbp]
	\centering
	\includegraphics[width=7cm]{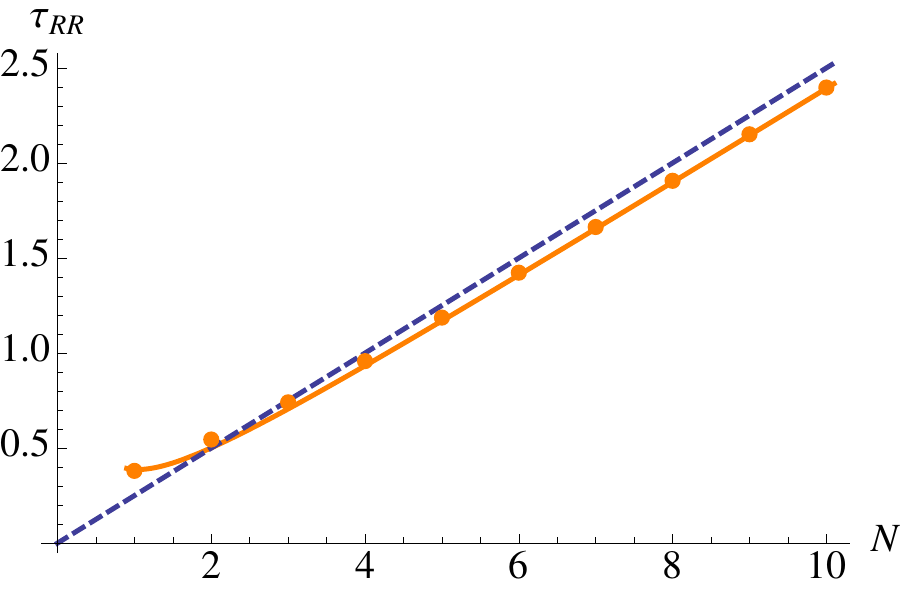}	\qquad
	\includegraphics[width=7cm]{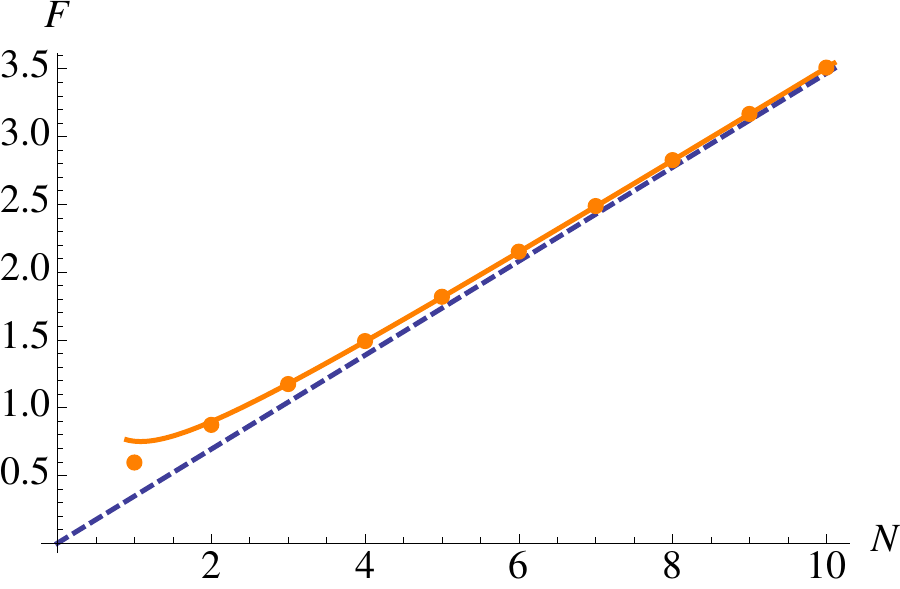}
	\caption{A plot of $\tau_{RR}$ as a function of $N$ for the Wess-Zumino model \eqref{eq:WZsuperpot} [Left]. A plot of $F$ for the same model [Right].
	The solid orange curves are calculated using the large-$N$ expansion at the fixed point with Eqs.\,\eqref{eq:WZlargeN} and \eqref{WZlargeNF}. The orange dots are computed  numerically. The dashed blue lines are the values at the IR free theory with Eqs.\,\eqref{WZIR} and \eqref{WZIRF}. }
	\label{fig:tauWZ}
\end{figure}

Numerical results for small $N$ in Table \ref{tab:WZtauRR} and Figure~\ref{fig:tauWZ} show that $\tau_{RR}^{XZ}$ is larger and smaller than $\tau_{RR}^Z$ for $N\le 2$ and $N\ge 3$, respectively. 
On the other hand, the free energy $F$ is always decreasing, consistent with the $F$-theorem. 
Therefore, we conclude that these are counter examples to the conjectured $C_T$-theorem.

\begin{table}[t]
\centering
	\begin{tabular}{c|cccccccccc
	}
		\hline
		$N$ & 1 & 2 & 3 & 4 & 5 & 6 & 7 & 8 & 9 & 10
		\\
		\hline\hline
		$\D_{Z}$ & 0.708 & 0.667 & 0.632 & 0.605 & 0.586 & 0.572 & 0.562 & 0.554  &0.548& 0.543
		\\
		$\tau_{RR}$ & 0.380 & 0.545 & 0.741 & 0.957 & 1.187 & 1.423 & 1.663 & 1.906  &2.151& 2.397
		\\
		$F$ & 0.595 & 0.872 & 1.174 & 1.491 & 1.817 & 2.150 & 2.487  &2.826 & 3.166 & 3.508
		\\
		\hline
	\end{tabular}
	
		\caption{The values of $\D_Z$, $\tau_{RR}$ and $F$ for the fixed points of the Wess-Zumino model with the superpotential \eqref{eq:WZsuperpot}.}
	\label{tab:WZtauRR}
\end{table}

%%%%%%%%%%%%%%%%%%%%%%%%%%%%%%%%%%%%%%%%%%%%%%%%%%%%%%%%%%%%%%%%%%%%%%%%%
%%%%%%%%%%%%%%%%%%%%%%%%%%%%%%%%%%%%%%%%%%%%%%%%%%%%%%%%%%%%%%%%%%%%%%%%%
\newpage
\vspace{1.3cm}
\centerline{\bf Acknowledgements}
We would like to thank F.\,Benini, S.\,Giombi, I.\,Klebanov,  B.\,Safdi, I.\,Yaakov for valuable discussions.
The work of T.\,N. was supported in part by the US NSF under Grants No.\,PHY-0844827 and
PHY-0756966. The work of K.\,Y. is supported in part by NSF grant PHY-0969448.
%%%%%%%%%%%%%%%%%%%%

\appendix
%%%%%%%%%%%%%%%%%%%%%%%%%%%%%%%%%%%%%%%%%%%%
\section{Hyperbolic gamma function}\label{sc:A}
%%%%%%%%%%%%%%%%%%%%%%%%%%%%%%%%%%%%%%%%%%%%
The hyperbolic gamma function is defined in the integral form \cite{van2007hyperbolic}:
\begin{align}\label{HypGamma}
	\Gamma_h \left(z; \xi_1, \xi_2 \right) = \exp \left[ i \int_0^\infty \frac{dx}{x} \left( \frac{z-\xi}{\xi_1 \xi_2 x} - \frac{\sin(2x(z-\xi))}{2\sin (\xi_1 x) \sin (\xi_2 x)}\right)\right] \ , 
\end{align}
for $z\in \BC$ satisfying $0 < \text{Im} (z) < \text{Im} (2\xi)$ where $\xi = (\xi_1 + \xi_2)/2$.
For brevity, we use a  notation $\Gamma_h[z] \equiv \Gamma_h(z; \xi_1, \xi_2)$.

A useful identity for our purpose is
\begin{align}\label{HGID}
	\frac{1}{\Gamma_h[z] \Gamma_h[-z]} &= -4\sin\left(\frac{\pi z}{\xi_1}\right) \sin\left(\frac{\pi z}{\xi_2}\right) \ . 
\end{align}

On a round sphere, \ie, $b=1$, the hyperbolic gamma function is written in terms of Jafferis's $\ell$-function \cite{Jafferis:2010un}:
\begin{align}
	\Gamma_h (z; i,i) = e^{\ell(1+i z)} \ ,
\end{align}
where
\begin{align}\label{Jafferisl}
	\ell(z) = -z\,\log(1-e^{2\pi i z}) + \frac{i}{2} \left( \pi z^2 + \frac{\text{Li}_2(e^{2\pi i z})}{\pi} \right) - \frac{i\pi}{12} \ .
\end{align}

%%%%%%%%%%%%%%%%%%%%%%%%%%%%%%%%%%%%%%%%%%%%
\section{Large-$N_f$ expansion for a general class of theories}\label{ap:B}
%%%%%%%%%%%%%%%%%%%%%%%%%%%%%%%%%%%%%%%%%%%%
Let us consider an $\CN=2$ theory with gauge group $G=\otimes_{A} G_A$, where $A$ labels each simple or $U(1)$ gauge group $G_A$.
We introduce chiral matter fields $\Phi_I$ labeled by $I$ in an irreducible representation $\CR_I=\otimes_{A} \CR_{I,A}$ of the gauge group,
and the $R$-charge of $\Phi_I$ will be denoted as $\Delta_I=\frac{1}{2}-a_I$. 
We introduce $N_I$ flavors of $\Phi_I$, and consider the limit where each $N_I $ becomes large with the ratio $N_I/N_J$ fixed.
We call this limit a large-$N_f$ limit, where $N_f$ represents the order of  flavor numbers $N_I$,
\ie, $N_I \sim O(N_f)$.

The partition function of the theory on a squashed $S^3$ is given by
\begin{align}
Z(b)= \int \prod_A [\CD \sigma_A]_b \prod_I (Z_{I}(b))^{N_I}  \ ,
\end{align}
where we have defined
\begin{align}
Z_{I}(b)= \prod_{\rho\in \CR_I} \Gamma_h \left[ \omega ( \rho(\sigma) + i \D_I)\right] \ ,
\end{align}
and
\begin{align}
\ [\CD  \sigma_A]_b= \frac{1}{\Vol(G_A)}  \cdot  d^{\dim G_A}  \sigma_A \cdot \exp (-i\pi k_A \omega^2\Tr \sigma_A^2) \cdot 
\prod_{\alpha_A \in Ad(G_A)} \left(\alpha_A(\sigma_A) \Gamma_h\left[ \omega \alpha_A (\sigma_A) \right] \right)^{-1} \ .
\end{align}
Note that the integral $d^{\dim G_A} \sigma_A$ is over all the $\dim G$ components of $\sigma_A$ (\ie, not only Cartan subalgebra),
and the product $\prod_{\alpha \in Ad(G_A)}$ is taken over all the roots and vanishing weights of the adjoint representation.
It is possible to reduce the integral $d^{\dim G_A} \sigma_A$ to an integral over the Cartan subalgebra by gauge-fixing as is in Eq.\,(\ref{PFb}), 
but we will find it more convenient to use the above expression in the following calculation.

\subsection{$F$-maximization for theories without superpotential}

For a round sphere $b=1$, $Z_{I}(b)$ and $[\CD \sigma_A]_b$ are simplified to 
\begin{align}\label{ZRI}
Z_I= \prod_{\rho\in \CR_I} e^{\ell (1-\Delta_I+i \rho_I) } \ ,
\end{align}
and
\begin{align}
\ [\CD  \sigma_A]=\frac{(2\pi)^{\dim G_A}}{\Vol(G_A)}  \cdot  d^{\dim G_A}  \sigma_A \cdot \exp (-i\pi k_A \Tr \sigma_A^2) \cdot 
\prod_{\alpha_A \in Ad(G_A)} \left( \frac{\sinh(\pi \alpha_A)}{\pi \alpha_A} \right) \ ,
\end{align}
where we have used abbreviation $\rho_I=\rho_I(\sigma)$ and $\alpha_A=\alpha_A(\sigma_A)$ for brevity.

We use saddle point approximation below. To this end, we expand $\ell (1-\Delta_I+i \rho_I)$ in a power series of $\rho_I$.
From the definition of $\rho_I$, the following identity holds
\begin{align}
\sum_{\rho\in \CR_I} (\rho_I)^k=\Tr_{\CR_I} \sigma^k \ .
\end{align}
It follows that we can rewrite the logarithm of Eq.\,\eqref{ZRI} as 
\begin{align}
\sum_{\rho\in \CR_I} \ell (1-\Delta_I+i \rho_I)=\sum_{k=0}^\infty \frac{i^k}{k !}\, \ell^{(k)}(1-\Delta_I) \Tr_{\CR_I} \sigma^k \ ,
\end{align}
where $\ell^{(k)}(z)$ is the $k$-th derivative of $\ell(z)$.
Similarly, we expand a part of the measure
\begin{align}
\prod_{\alpha \in Ad} \left( \frac{\sinh(\pi \alpha_A)}{\pi \alpha_A} \right)=1+\frac{\pi^2}{6}\Tr_{Ad(G_A)} \sigma_A^2+O(\sigma_A^4) \ .
\end{align}

We will make the following assumption for simplicity. If the gauge group $G$ contains $U(1)$ (and $U(1)' $) gauge group(s), let $q$ ($q'$) be the
$U(1)$ ($U(1)')$ charges of matter fields. Then, we assume~\footnote{In terms of Feynman diagrams, Eq.\,(\ref{eq:cond1}) forbids 1-loop tadpole diagrams 
of a $U(1)$ multiplet, while Eq.\,(\ref{eq:cond2}) forbids 1-loop kinetic mixing diagrams between two $U(1)$ multiplets.
}
\begin{align}
\sum_{\text{all matters}} q &=0 \ ,  \label{eq:cond1}  \\
\sum_{\text{all matters}} q q' &=0 \ . \label{eq:cond2}
\end{align}
See section \ref{ssc:Chiral} for an example which does not satisfy Eq.\,(\ref{eq:cond1}).

The matrix $\sigma$ in the representation $\CR_I$ is
\begin{align}
\sigma _{\CR_I} = \sum_A  1_{\CR_{I,1}} \otimes  \cdots \otimes 1_{\CR_{I,A-1}} \otimes \sigma_{\CR_{I,A}} \otimes 1_{\CR_{I,A+1}} \otimes \cdots \ ,
\end{align}
where $1_{\CR_{I,A}}$ is the unit matrix in the representation $\CR_{I,A}$ and $\sigma_{\CR_{I,A}}$ is the matrix $\sigma_A$ in $\CR_{I,A}$.
The assumptions (\ref{eq:cond1}) and (\ref{eq:cond2}) leads to 
\begin{align}
\sum_I N_I \Tr_{\CR_I} \sigma&=0 \ , \\
\sum_I N_I \Tr_{\CR_I} \sigma^2 &= \sum_I N_I 
\left [ \sum_A \frac{\dim \CR_I}{\dim \CR_{I,A}} \Tr_{\CR_{I, A}} \sigma_A^2 + \sum_{A \neq B} \frac{\dim \CR_I}{\dim \CR_{I,A} \dim \CR_{I,B}}
(\Tr_{\CR_{I, A}} \sigma_A)( \Tr_{\CR_{I, B}} \sigma_B)
\right]
  \nonumber \\
&=  \sum_I N_I \sum_A \frac{ \dim \CR_I}{\dim \CR_{I, A}} t_{\CR_{I, A}}  \Tr  \sigma_A^2  \nonumber \\
&\equiv \sum_I N_I \dim \CR_I \sum_A \hat{t}_{\CR_{I, A}}\Tr \sigma_A^2 \ .
\end{align}
Here $t_{\CR_{I, A}}$ is the Dynkin index of the representation $\CR_{I, A}$ defined as
 $\Tr_{\CR_{I, A}} (T_a T_b)=t_{\CR_{I, A}} \delta_{ab}$, where gauge group generators $T_a$ are assumed to be normalized as $\Tr (T_a T_b)=\delta_{ab}$.
 We have also defined $\hat{t}_{\CR_{I, A}} ={t}_{\CR_{I, A}}/\dim \CR_{I, A}$ for simplicity.
 
As we will see, $a_I=1/2-\Delta_I$ is of order $1/N_f$. Then, up to the first subleading corrections to the partition function, we obtain
\begin{align}
Z&=\text{const.} \int \prod_{A} d^{\dim G_A}  \sigma_A \cdot 
\exp \left[ -\sum_A \left( \frac{\pi^2}{4}\sum_I \hat{t}_{\CR_{I, A}} N_I \dim \CR_I  + i\pi k_A \right) \Tr \sigma_A^2 \right] \nonumber \\
& \cdot \Bigg( 1+\frac{\pi^2}{4}\sum_I a_I^2 N_I \dim \CR_I -\pi^2 \sum_I \sum_A a_I N_I \dim \CR_I  \hat{t}_{\CR_{I, A}} \Tr \sigma_A^2 \nonumber \\
&~~~~~+(a_I\text{-independent terms})+O(N_f^{-2})\Bigg) \ ,
\end{align}
where we have used $a_I \sim O(N_f^{-1})$ and $\sigma^{k} \sim O(N_f^{-k/2})$.
We have also used $\ell'(1/2)=0,~\ell''(1/2)=\pi^2/2$ and $\ell'''(1/2)=2\pi^2$.

The above integral is just a gaussian integral and can be done easily.
From the $F$-maximization $\partial \Re [ \log Z] /\partial a_I=0$, we obtain
\begin{align}
a_I= \frac{4}{\pi^2} \sum_A \frac{\dim G_A}{1+\kappa_A^2} \left( \frac{\hat{t}_{\CR_{I, A}} }{ \sum_J \hat{t}_{\CR_{J, A}} N_J \dim \CR_J   } \right) +O(N_f^{-2}) 
\label{eq:generalRcharge} \ ,
\end{align}
where we have defined 
\begin{align}
\kappa_A= \frac{4k_A}{\pi \sum_J \hat{t}_{\CR_{J, A}} N_J \dim \CR_J  } \ . \label{eq:kappaA}
\end{align}

\subsection{Computation of $\tau_{RR}$}
Next we discuss the computation of $\tau_{RR}$. First, we define the expectation value of a function $f(\sigma)$ as
\begin{align}
\vev{f(\sigma)} = \frac{\int \prod_A [\CD \sigma_A] \prod_I (Z_{\CR_I})^{N_I} f(\sigma)  }{ \int \prod_A [\CD \sigma_A] \prod_I (Z_{\CR_I})^{N_I} }  \ .
\end{align}
Using this notation, $\tau_{RR}$ is given by
\begin{align}
\tau_{RR} 
&=\Re \vev{\sum_I N_I f_{\CR_I}(1-\Delta_I,\sigma)+\sum_A g_A(\sigma) } \ ,
\end{align}
where we have defined
\begin{align}
f_{\CR_I}(z,\sigma)=&\frac{2}{\pi^2} \sum_{\rho\in \CR_I} \int^\infty_0 dx \Bigg[ 
z \left( \frac{1}{x^2}-\frac{\cosh (2x(z+i \rho_I))}{\sinh^2 (x)} \right) 
+ \frac{( \sinh (2x)-2x) \sinh(2x(z+i\rho_I))}{2\sinh^4(x)}    \Bigg] \ , \\
g_A(\sigma)=&-\frac{1}{\pi^2} \sum_{\alpha_A \in Ad(G_A)} \frac{(\pi \alpha_A ) \sinh (2 \pi \alpha_A) -2(\pi \alpha_A)^2}{\sinh^2(\pi \alpha_A)} \ .
\end{align}

One can check that $\vev{g_A(\sigma)}$ starts from the order $N^{-1}_f$ and we neglect it in this appendix.
On the other hand, by expanding $f_{\CR_I}(1-\Delta_I,\sigma)$ in terms of $a_I$ and $\sigma$, some computation yields
\begin{align}
f_{\CR_I}(1-\Delta_I,\sigma)&=\dim \CR_I \left( \frac{1}{4} +\frac{1}{3} a_I \right)+\left(2 -\frac{\pi^2}{4} \right) \Tr_{\CR_I } \sigma^2+O(\sigma^4) \ .
\end{align}
In the saddle point approximation used in the previous subsection, we obtain
\begin{align}
\vev{\sum_I N_I \Tr_{\CR_I } \sigma^2}=
\frac{2}{\pi^2} \sum_A \frac{\dim G_A}{1-i\kappa_A} 
+O(N_f^{-1}) \ .
\end{align}
Therefore, $\tau_{RR}$ is given by
\begin{align}
\tau_{RR} &=\frac{1}{4}\left [ \sum_I \left(1 +\frac{4}{3}a_I\right)N_I \dim \CR_I +\frac{2(8 - \pi^2)}{ \pi^2}  \sum_A \frac{\dim G_A}{1+\kappa_A^2} \right]
+O(N_f^{-1}) \ .
\end{align}
For theories without superpotential, the $R$-charges are determined in Eq.\,(\ref{eq:generalRcharge}).
In this case, $\tau_{RR}$ is given by
\begin{align}
\tau_{RR}&=\frac{1}{4}\left [ \sum_I N_I \dim \CR_I + \frac{2(32-3\pi^2)}{ 3\pi^2}  \sum_A \frac{\dim G_A}{1+\kappa_A^2} \right] 
+O(N_f^{-1})\ .
\end{align}

%%%%%%%%%%%%%%%%%%%%%%%%%%%%%%%%%%%%%%%%%%%%%%%%%%%%%%%%%
%%%%%%%%%%%%%%%%%%%%%%%%%%%%%%%%%%%%%%%%%%%%%%%%%%%%%%%%%

\bibliographystyle{JHEP}
\bibliography{tauRG}

\end{document}